\documentclass[aps,prd,twocolumn,groupedaddress]{revtex4-1}
\usepackage{amssymb}
\usepackage{graphicx}
\usepackage{dcolumn}
\usepackage{bm}
\usepackage{hyperref}
\begin{document}
\title{Observational tests of Galileon gravity with growth rate}
\author{Koichi Hirano}
\email[]{hirano@ichinoseki.ac.jp}
\affiliation{Department of Physics, Ichinoseki National College of Technology, Ichinoseki 021-8511, Japan}
\author{Zen Komiya}
%\email{zen@komiyake.com}
\affiliation{Department of Physics, Tokyo University of Science, Tokyo 162-8601, Japan}
\date{\today}
\begin{abstract}
We compare observational data of growth rate with the prediction by Galileon theory. For the same value of the energy density parameter $\Omega_{m,0}$, the growth rate in Galileon models is enhanced compared with the $\Lambda$CDM case, due to the enhancement of Newtonfs constant. The smaller $\Omega_{m,0}$ is, the more suppressed growth rate is. Hence the best fit value of $\Omega_{m,0}$ in the Galileon model is 0.16 from only the growth rate data, which is considerably smaller than such value obtained from observations of supernovae Ia, the cosmic microwave background and baryon acoustic oscillations. This result seems to be qualitatively the same in most of the generalized Galileon models. We also find the upper limit of the Brans--Dicke parameter to be $\omega < -40$, from the growth rate data. More and better growth rate data are required to distinguish between dark energy and modified gravity.
\end{abstract}
\maketitle
\section{INTRODUCTION}
Cosmological observations, such as type Ia supernovae (SNIa) \cite{rie1998,per1999} and the cosmic microwave background (CMB) anisotropies \cite{kom2010}, indicate that the universe is undergoing an accelerated phase of expansion. This late-time acceleration is one of the biggest mysteries in modern cosmology. The conventional explanation is that it is caused by the cosmological constant or dark energy \cite{rat1988,cal2002,kom2006,kom2005}. This would mean that the universe is mostly filled with an unknown energy-momentum component. The cosmological constant is the standard candidate for dark energy. To explain the current acceleration of the universe, the cosmological constant must have an incredibly small value. However, its value cannot be explained by current particle physics and it is affected by fine-tuning problems and the coincidence problem.

An alternative explanation for the current accelerated expansion of the universe is to extend general relativity to a more general theory of gravity at long distances. Several modified gravity approaches have been proposed including $f(R)$ gravity \cite{fel2010}, scalar-tensor theories \cite{hir2010c,hir2008a,har2008}, and the Dvali--Gabadazde--Porrati (DGP) braneworld model \cite{dva2000,hir2010a,hir2011a}. The DGP model, however, is plagued by the ghost problem \cite{koy2007} and is incompatible with cosmological observations \cite{xia2009}.

As an alternative to general relativity, Galileon gravity models have recently been proposed \cite{nic2009,cho2009,sil2009,kob2010a,kob2010b,gan2010,fel2010a,fel2010b,def2010}. These models are constructed by introducing a scalar field with a self-interaction whose Lagrangian is invariant under Galileon symmetry $\partial_\mu\phi\rightarrow\partial_\mu\phi+b_\mu$, which maintains the equation of motion as a second-order differential equation. This prevents the theory from exhibiting a new degree of freedom, and perturbation of the theory does not lead to ghost or instability problems. The simplest term for the self-interaction is $\Box\phi(\nabla\phi)^2$, which appears in the 4-dimensional effective theory of the DGP model. The self-interaction term $\Box\phi(\nabla\phi)^2$ induces decoupling of the Galileon field $\phi$ from gravity at small scales by the so-called Vainshtein mechanism \cite{vai1972}. This allows the Galileon theory to recover general relativity at scales around the high density region, which is consistent with solar-system experiments.

Galileon theory has been covariantized and considered in curved backgrounds
\cite{def2009a,def2009b}. It has been shown that Galileon symmetry cannot be preserved once the theory is covariantized; however, it is possible to keep the equation of motion as a second-order differential equation, that is, free from ghostlike instabilities. Galileon gravity can induce self-accelerated expansion of the late-time universe. Hence, inflation models inspired by Galileon theory have been proposed \cite{kob2010c,miz2010,kam2010}. In Ref. \cite{nes2010}, the parameters of the generalized Galileon cosmology were constrained by observational data of supernovae Ia (SN Ia), the cosmic microwave background (CMB) and baryon acoustic oscillations (BAO). The evolution of matter density perturbations in Galileon models has also been studied \cite{kim2010,fel2010c,sil2009,kob2010a,kob2010b}.

In this letter, we compare observational data of growth rate with the predictions of Galileon theory. Even though the background expansion history in modified gravity is almost identical to that of the standard $\Lambda$CDM model or dark energy models, the evolution of matter density perturbations in modified gravity is different from that of the $\Lambda$CDM model or dark energy models. Thus it is important to study the growth history of perturbations in order to distinguish modified gravity from models based on the cosmological constant or dark energy. Hence, we compute the growth rate of matter density perturbations in Galileon cosmology and compare it with observational data.
%This is a powerful discriminant among models of cosmic acceleration.

This letter is organized as follows. In the next section, we describe the background evolution and the evolution of linear perturbations in Galileon cosmology. In Sec. \ref{obs} we study the observational tests of Galileon gravity with growth rate. Finally, a summary is given in Sec. \ref{con}.

\section{GALILEON GRAVITY MODEL \label{mod}}
\subsection{BACKGROUND EVOLUTION}
The action we consider is given by \cite{sil2009,kob2010a}
\begin{equation}
S=\int{d^4x\sqrt{-g}\left[\phi R-\frac{\omega}{\phi}(\nabla\phi)^2+f(\phi)\Box\phi(\nabla\phi)^2+L_m\right]},
\end{equation}
where $(\nabla\phi)^2=g^{\mu\nu}\nabla_\mu\phi\nabla_\nu\phi$, $\Box\phi=g^{\mu\nu}\nabla_\mu\nabla_\nu\phi$, $L_m$ is the matter Lagrangian. $\omega$ is the Brans--Dicke parameter and $f(\phi)$ is a function of the Galileon field $\phi$. Variation with respect to the metric gives the Einstein equations, and variation with respect to the Galileon field $\phi$ leads to the equation of motion. For Friedmann--Robertson--Walker spacetime, the Einstein equations give
\vspace{0.1mm}
\begin{widetext}
\begin{equation}
3H^2=\frac{\rho}{2\phi}-3HP+\frac{\omega}{2}P^2+\phi^2f(\phi)\left(3H-\frac{d_1}{2}P\right)P^3, \label{freq1}
\end{equation}
\begin{equation}
-3H^2-2\dot{H}=\frac{p}{2\phi}+\dot{P}+P^2+2HP+\frac{\omega}{2}P^2-\phi^2f(\phi)\left(\dot{P}+\frac{d_1+2}{2}P^2\right)P^2 \label{freq2}
\end{equation}
and the equation of motion for the Galileon field gives
\begin{equation}
6\left(2H^2+\dot{H}\right)-\omega\left(2\dot{P}+P^2+6HP\right)-\phi^2f(\phi)\left[6\left(2H\dot{P}+\dot{H}P+2HP^2+3H^2P\right)P-4d_1P^2\dot{P}-\left({d_1}^2+3d_1+d_2\right)P^4\right]=0,
\end{equation}
\end{widetext}
where an overdot represents differentiation with respect to cosmic time $t$, $H=\dot{a}/a$ is the Hubble expansion rate, $P\equiv\dot{\phi}/\phi$, and $d_n={\rm d}^n\ln{f(\phi)}/{\rm d}\ln{\phi}^n$. $\rho$ is the energy density and $p$ is the pressure. In this letter, we consider a spatially flat Universe ($k = 0$) only.

The Friedmann equation (\ref{freq1}) can be written in the form
\begin{equation}
3H^2=\frac{1}{M_{\rm pl}^2}(\rho+\rho_\phi),
\end{equation}
where the effective dark energy density $\rho_\phi$ is defined as
\begin{eqnarray}
\rho_\phi & = & 2\phi\left[-3HP+\frac{\omega}{2}P^2+\phi^2f(\phi)\left(3H-\frac{d_1}{2}P\right)P^3\right] \nonumber \\
& & +3H^2\left(M_{\rm pl}^2-2\phi\right).
\end{eqnarray}

In the numerical analysis in this letter, since we are interested in the basic parameters of Galileon theory, we consider a specific model with
\begin{equation}
f(\phi)=\frac{r_c^2}{\phi^2},
\end{equation}
where $r_c$ is the crossover scale \cite{cho2009}. The energy density parameter of matter is defined as $\Omega_{m}=\rho/3M_{\rm pl}^2H^2$.

At early times, we set $\phi\simeq 1/16\pi G$, and the Einstein equations (\ref{freq1}) and (\ref{freq2}) reduce to the usual forms: $3H^2 = 8\pi G\rho$ and $3H^2 + 2\dot{H}=-8\pi Gp$, that is, general relativity is recovered below a certain scale. At late times, in order to describe the cosmic acceleration today, the value of $r_c$ must be fine-tuned. 

\subsection{DENSITY PERTURBATIONS}
The evolution equation for the cold dark matter overdensity $\delta$ in linear theory is governed by
\begin{equation}
\ddot{\delta}+2H\dot{\delta}-4\pi G_{\rm eff}\rho\delta\simeq0,
\end{equation}
where $G_{\rm eff}$ represents the effective Newtonfs constant in Galileon gravity, which is obtained as
\begin{equation}
G_{\rm eff}=\frac{1}{16\pi\phi}\left[1+\frac{(1+f(\phi)\dot{\phi}^2)^2}{F}\right],
\end{equation}
where
\begin{equation}
F=3+2\omega+\phi^2f(\phi)\left[4\frac{\ddot{\phi}}{\phi}-2\frac{\dot{\phi}^2}{\phi^2}+8H\frac{\dot{\phi}}{\phi}-\phi^2f(\phi)\frac{\dot{\phi}^4}{\phi^4}\right].
\end{equation}
The effective Newtonfs constant $G_{\rm eff}$ is close to $G$ at early times but increases at late times. (The effective Newtonfs constant may depend on the scale $k$, but we are not concerned about this point as we do not consider the power spectrum of $\delta$ in this letter.)

\begin{figure}
\includegraphics[width=90mm]{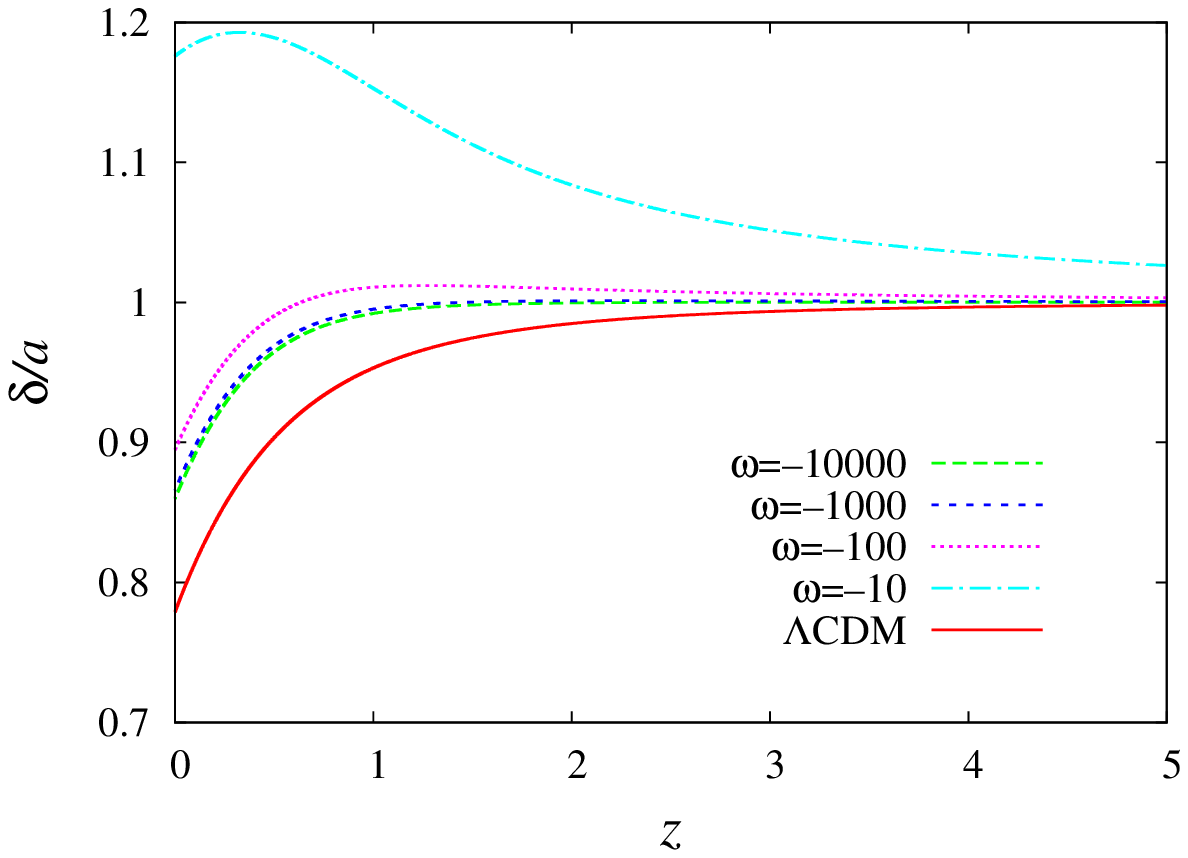}
\caption{Growth function $\delta/a$ in Galileon gravity as a function of redshift $z$ for various values of the Brans--Dicke parameter $\omega$. The parameters are given by $\Omega_{m,0}=0.30$. \label{g_w}}
%\end{figure}
\vspace{3.5mm}
%\begin{figure}[h!]
\includegraphics[width=90mm]{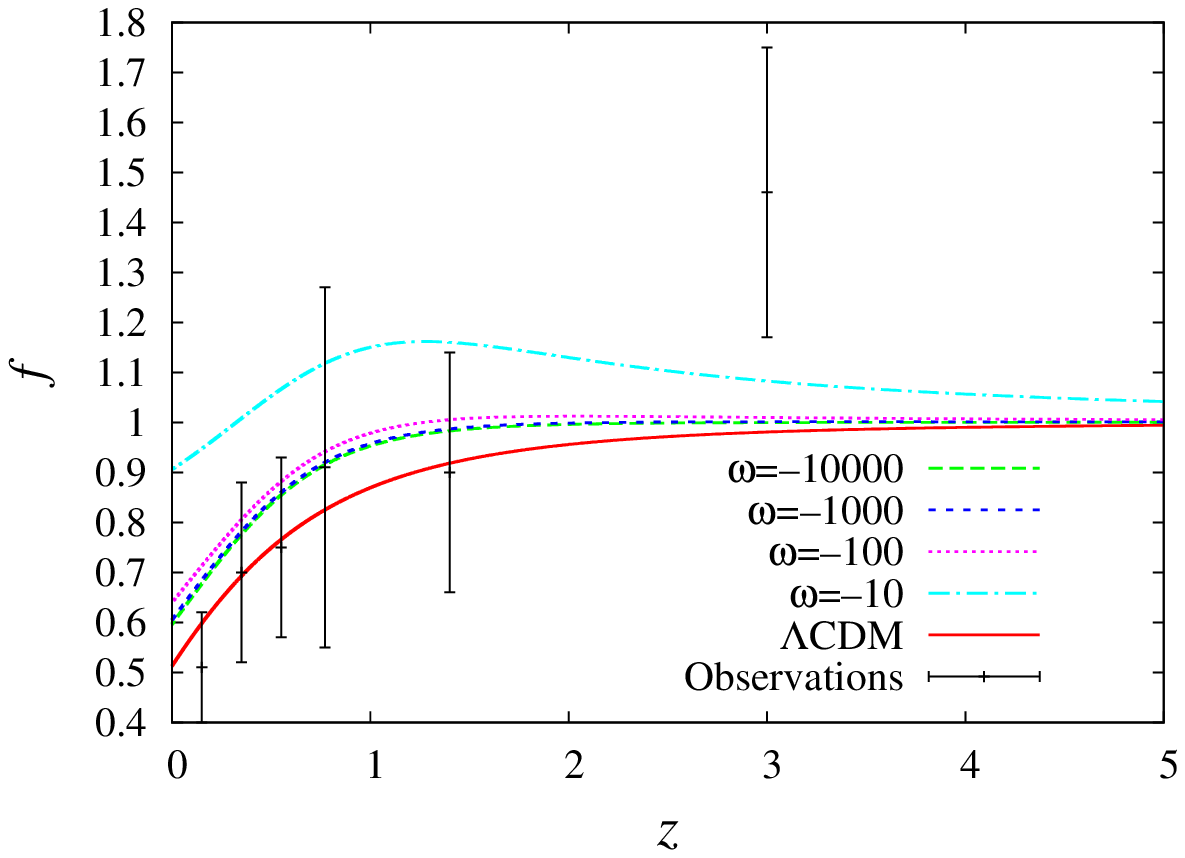}
\caption{Growth rate $f$ in Galileon gravity as a function of redshift $z$ for various values of the Brans--Dicke parameter $\omega$. The parameters are given by $\Omega_{m,0}=0.30$. \label{f_w}}
%\end{figure}
%
%\begin{figure}[bh!]
\includegraphics[width=90mm]{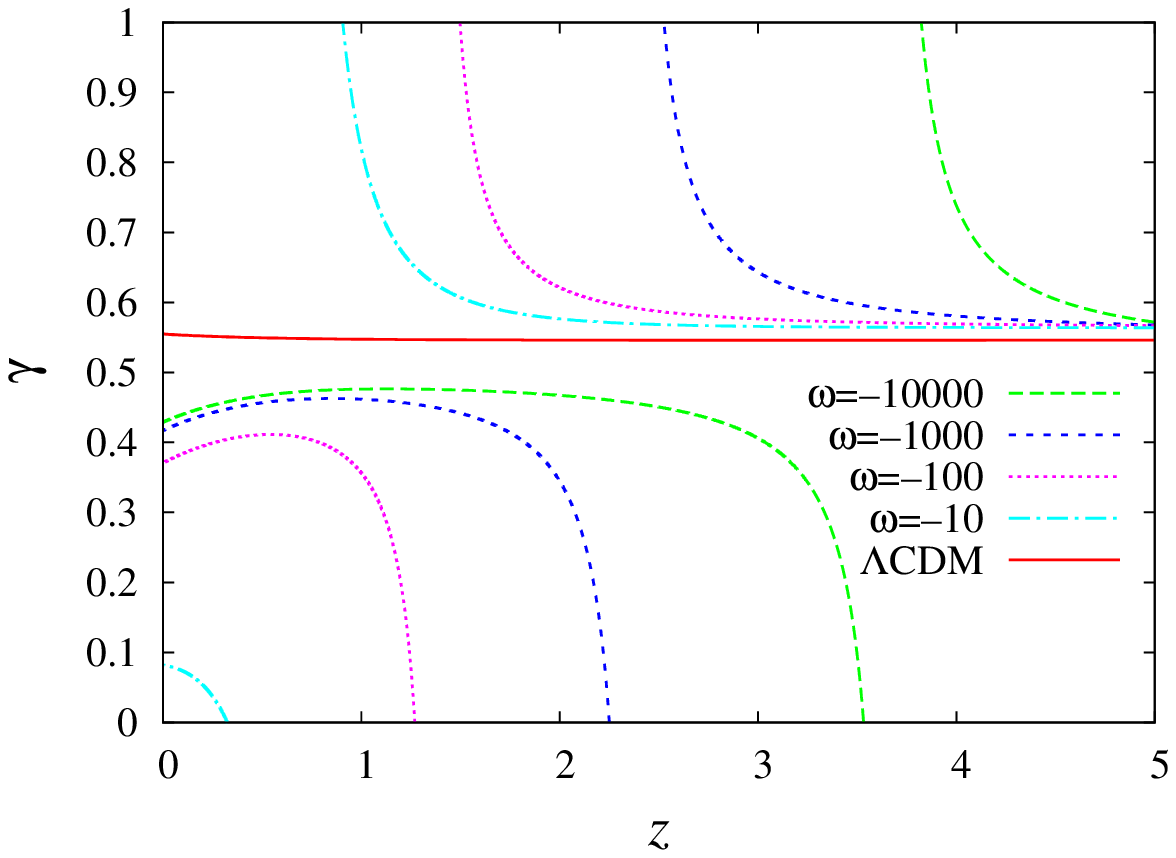}
\caption{Growth index $\gamma$ in Galileon gravity as a function of redshift $z$ for various values of the Brans--Dicke parameter $\omega$. The parameters are given by $\Omega_{m,0}=0.30$. \label{c_w}}
\end{figure}

\begin{figure}
\includegraphics[width=90mm]{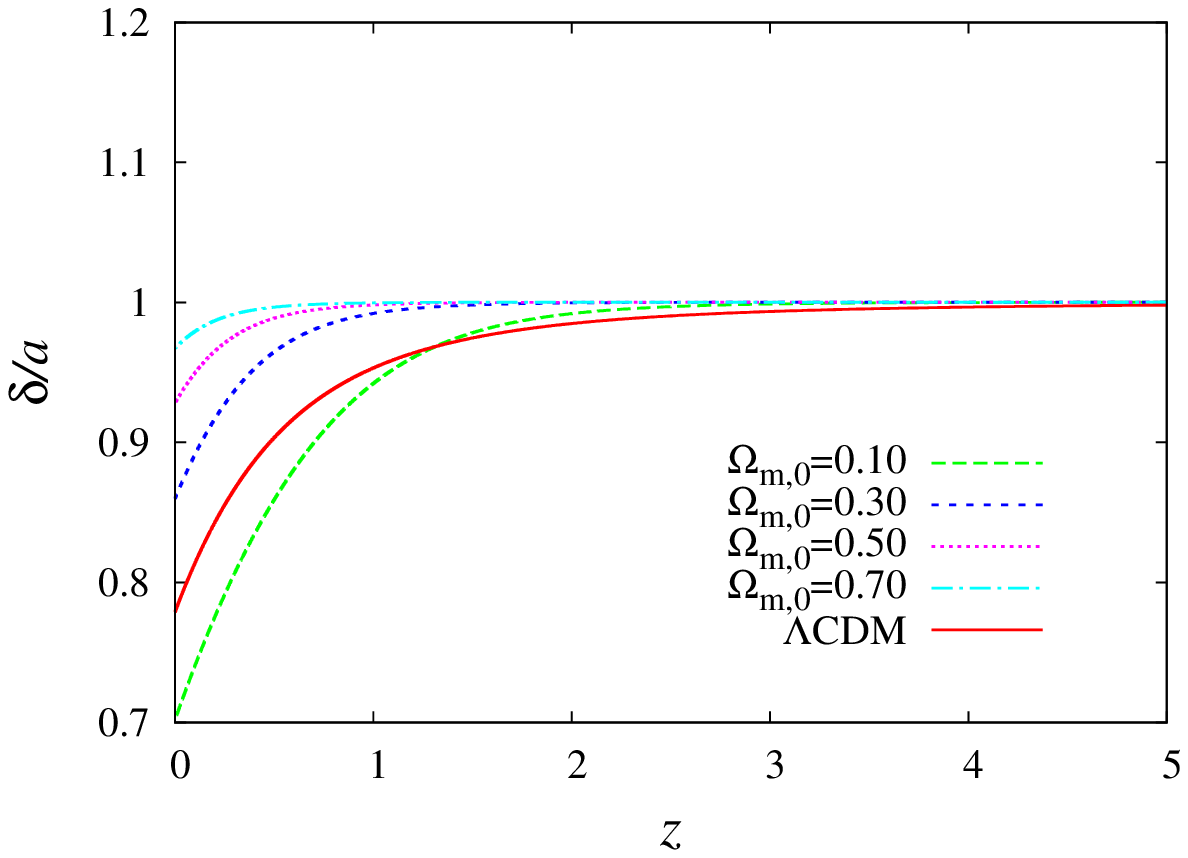}
\caption{Growth function $\delta/a$ in Galileon gravity as a function of redshift $z$ for various values of today's energy density parameter of matter $\Omega_{m,0}$. The parameters are given by $\omega=-10000$. \label{g_m}}
%\end{figure}
%
%\begin{figure}[h!]
\includegraphics[width=90mm]{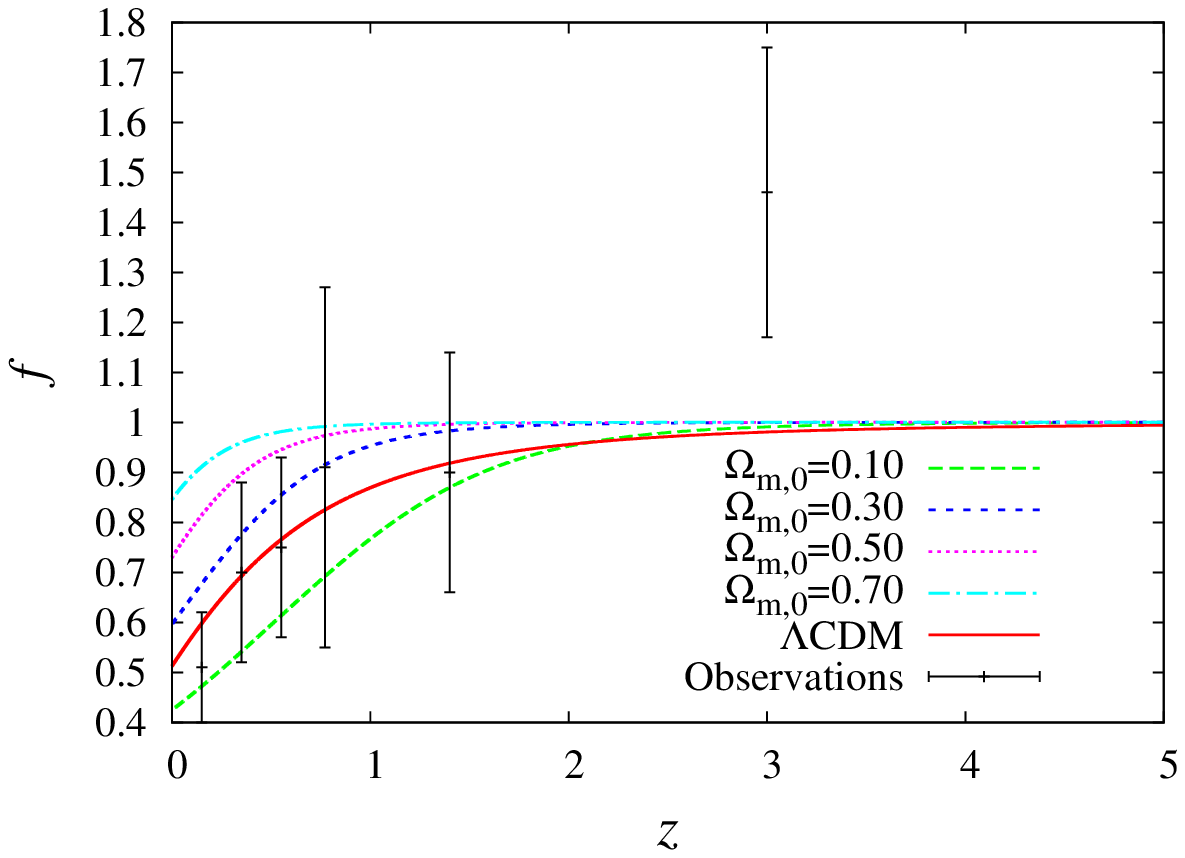}
\caption{Growth rate $f$ in Galileon gravity as a function of redshift $z$ for various values of today's energy density parameter of matter $\Omega_{m,0}$. The parameters are given by $\omega=-10000$. \label{f_m}}
%\end{figure}
%
%\begin{figure}[bh!]
\includegraphics[width=90mm]{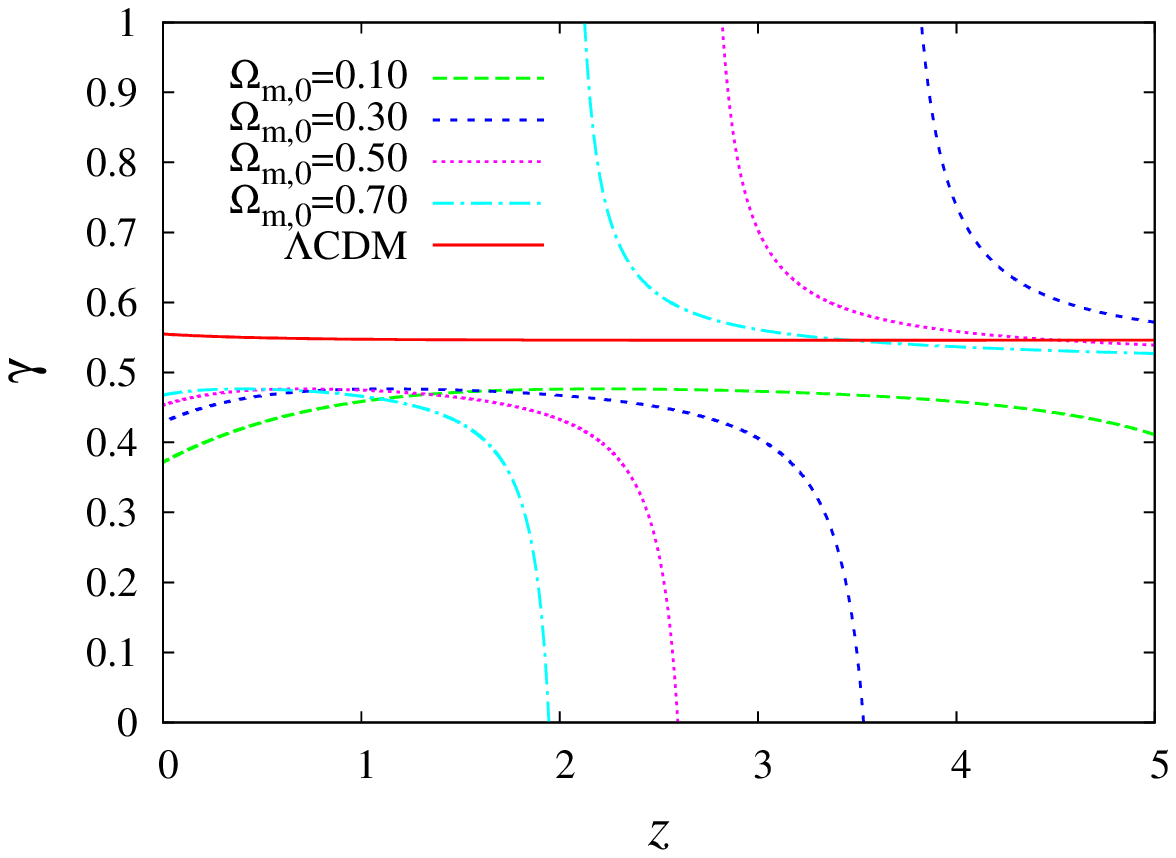}
\caption{Growth index $\gamma$ in Galileon gravity as a function of redshift $z$ for various values of today's energy density parameter of matter $\Omega_{m,0}$. The parameters are given by $\omega=-10000$. \label{c_m}}
\end{figure}

We set the initial condition $\delta\approx a$ at early times. Solving the evolution equation numerically we obtain the growth factor $\delta/a$ in Galileon gravity as shown in Fig. \ref{g_w} and Fig. \ref{g_m}. The linear growth rate is written as
\begin{equation}
f=\frac{{\rm d}\ln{\delta}}{{\rm d}\ln{a}}.
\end{equation}
The growth rate can be parameterized by the growth index $\gamma$ defined by
\begin{equation}
f=\Omega_m^\gamma.
\end{equation}

We plot the growth rate $f$ in Galileon gravity in Fig. \ref{f_w} and Fig. \ref{f_m}, and the growth index $\gamma$ in Galileon gravity in Fig. \ref{c_w} and Fig. \ref{c_m}. $\omega$ is the (constant) Brans--Dicke parameter and $\Omega_{m,0}$ is the energy density parameter of matter at the present day. We set $\Omega_{m,0}=0.30$ for the Galileon model in Fig. \ref{g_w}, \ref{f_w}, \ref{c_w} and $\omega=-10000$ for the Galileon model in Fig. \ref{g_m}, \ref{f_m}, \ref{c_m}. $\Omega_{m,0}$ for the $\Lambda$CDM model is $0.30$ in these Figs.

For the same value of $\Omega_{m,0}$, the growth rate $f$ in Galileon models is enhanced compared with the $\Lambda$CDM case, due to the enhancement of Newtonfs constant. The smaller $\Omega_{m,0}$ is, the more suppressed the growth rate is. We describe the observational data of Fig. \ref{f_w} and Fig. \ref{f_m} in the next section.

\section{OBSERVATIONAL TESTS \label{obs}}
\subsection{Observational data}
In Table \ref{fdata} we list the growth rate data used in our analysis: the linear growth rate $f\equiv{\rm d}\ln{\delta}/{\rm d}\ln{a}$ from a galaxy power spectrum at low redshifts \cite{haw2003,ver2002,teg2006,ros2006,guz2008,ang2006} and a Lyman-$\alpha$ growth factor measurement obtained with the Lyman-$\alpha$ power spectrum at z = 3 \cite{mac2005}.

\begin{table}[h!]
\caption{Currently available data for linear growth rates $f_{obs}$ used in our analysis. $z$ is redshift and $\sigma$ is the 1$\sigma$ uncertainty of the growth rate data. \label{fdata}}
\begin{ruledtabular}
\begin{tabular}{c c c c c}
$z$ & $f_{obs}$ & $\sigma$ & Ref. \\
\hline
0.15 & 0.51 & 0.11 & \cite{haw2003,ver2002} \\
0.35 & 0.70 & 0.18 & \cite{teg2006} \\
0.55 & 0.75 & 0.18 & \cite{ros2006} \\
0.77 & 0.91 & 0.36 & \cite{guz2008} \\
1.40 & 0.90 & 0.24 & \cite{ang2006} \\
3.00 & 1.46 & 0.29 & \cite{mac2005} \\
\end{tabular}
\end{ruledtabular}
\end{table}

The corresponding $\chi^2$ is given by:
\begin{equation}
\chi^2=\sum_{i=1}^6\frac{(f(z_i)-f_{obs}(z_i))^2}{\sigma(z_i)^2}. \label{chi2}
\end{equation}

To determine the best value and the allowed region of the parameters, we minimize the $\chi^2$ and use the maximum likelihood method.

\subsection{Numerical results}
We now present our main results for the observational tests with growth rate.

\begin{figure}[h!]
\includegraphics[width=87mm]{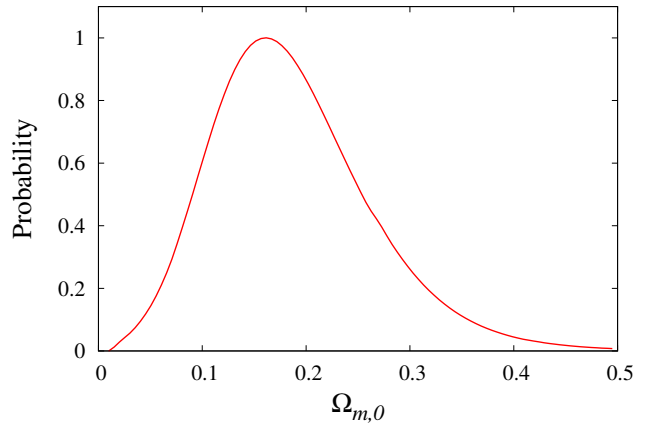}
\caption{1D probability distribution of the energy density parameter of matter $\Omega_{m,0}$ for the Galileon model from the growth rate data.\label{1domegam}}
\end{figure}

In Fig. \ref{1domegam}, we plot the probability distribution of $\Omega_{m,0}$ for the Galileon model from the growth rate data. The best fit value is $\Omega_{m,0}$ = 0.16. This is considerably smaller than such value obtained from observations of SNIa and CMB and BAO \cite{nes2010, kim2010}. We obtained the constraint as follows:

\begin{equation}
\Omega_{m,0}=0.16^{+0.18}_{-0.11}.~~~~~~~~(95\%~{\rm C.L.})
\end{equation}

\begin{figure}[h!]
\includegraphics[width=87mm]{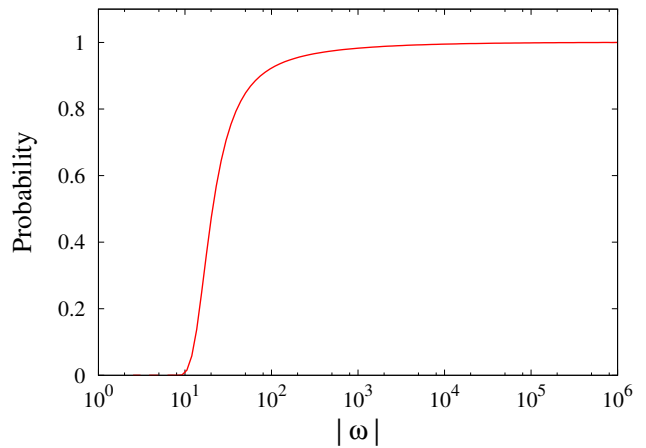}
\caption{1D probability distribution of absolute value of the Brans--Dicke parameter $|\omega |$ (note that $\omega < 0$) for the Galileon model from the growth rate data. \label{1dxi}}
\end{figure}

In Fig. \ref{1dxi}, we plot the probability distribution of the Brans--Dicke parameter $\omega$ for the Galileon model from the growth rate data. We obtained the constraint as follows:

\begin{equation}
\omega < -40.~~~~~~~~(95\%~{\rm C.L.})
\end{equation}

\begin{figure}
\includegraphics[width=89mm]{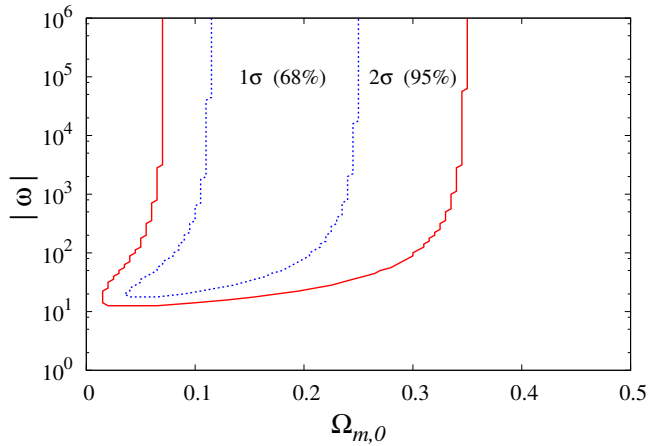}
\caption{Probability contours in the ($\Omega_{m,0}$, $|\omega |$)-plane for the Galileon model. The dotted (blue) and solid (red) contours show the 1$\sigma$ (68\%) and 2$\sigma$ (95\%) confidence limits, respectively, from the growth rate data. \label{2d}}
\end{figure}

In Fig. \ref{2d}, we plot the probability contours in the ($\Omega_{m,0}$, $|\omega |$)-plane for the Galileon model. The dotted (blue) and solid (red) contours show the 1$\sigma$ (68\%) and 2$\sigma$ (95\%) confidence limits, respectively, from the growth rate data. ($|\omega |$ is the absolute value of the Brans--Dicke parameter; $\omega < 0$.)

\begin{figure}
\includegraphics[width=89mm]{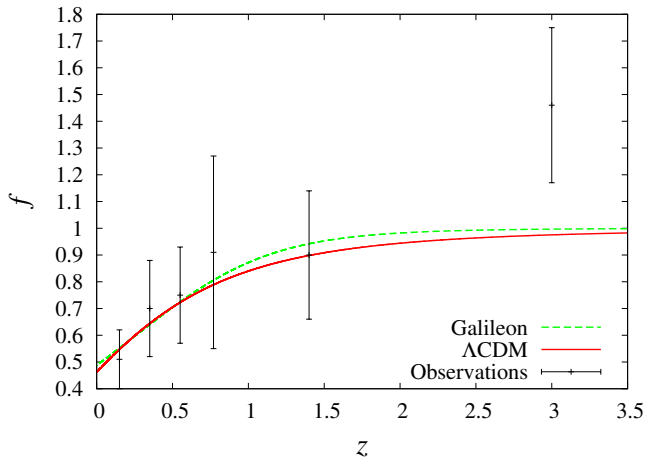}
\caption{Growth rate $f$ in best fit models of Galileon theory and $\Lambda$CDM with parameters in Table \ref{table2}. \label{best}}
\end{figure}

In Table \ref{table2}, we list the best fit parameters, the $\chi^2$ values (Eq. (\ref{chi2})), and the differences of the Akaike information criterion (AIC) \cite{aka1974} and the Bayesian information criterion (BIC) \cite{sch1978}, for the Galileon model and the $\Lambda$CDM model, from growth rate data.
The $\chi^2$ value for the Galileon model is smaller than that for the $\Lambda$CDM model. However, the values of AIC and BIC for the $\Lambda$CDM model are smaller than for the Galileon model, because there is one more free parameter in the Galileon model than in the $\Lambda$CDM model. In the $\Lambda$CDM model, the value of $\Omega_{m,0}$ from only the growth rate data is consistent with value obtained from observations of SNIa and CMB and BAO.

\begin{table}
\caption{Results of observational tests from the growth rate data. \label{table2}}
\begin{tabular}{l l l r r}
\hline\hline
Model  &  Best fit parameters  &  $\chi^2$  &  $\Delta$AIC  &  $\Delta$BIC \\
\hline
$\Lambda$CDM  &  $\Omega_{m,0}=0.25$  &  3.138  &  0.00  &  0.00 \\ 
Galileon ~ &  $\Omega_{m,0}=0.17$, $\omega=-2.0\times 10^6$ ~ &  2.943  & 1.81  & 1.60  \\
\hline\hline
\end{tabular}
\end{table}

In Fig. \ref{best}, we plot the growth rate $f$ in the best fit models of the Galileon theory and $\Lambda$CDM. More and better growth rate data are required to distinguish between dark energy and modified gravity.

\section{SUMMARY \label{con}}
For the same value of $\Omega_{m,0}$, the growth rate $f$ in Galileon models is enhanced compared with the $\Lambda$CDM case, due to the enhancement of Newtonfs constant. The smaller $\Omega_{m,0}$ is, the more suppressed growth rate is. Hence the best fit value of $\Omega_{m,0}$ in the Galileon model from only the growth rate data is $\Omega_{m,0}$=0.16. This is considerably smaller than such value obtained from observations of SNIa and CMB and BAO \cite{nes2010, kim2010}. This result seems to be qualitatively the same in most of the generalized Galileon models. On the other hand, in the $\Lambda$CDM model, the value of $\Omega_{m,0}$ from only the growth rate data is consistent with value obtained from observations of SNIa and CMB and BAO. We also find the upper limit of the Brans--Dicke parameter to be $\omega < -40$, from the growth rate data. More and better growth rate data are required to distinguish between dark energy and modified gravity.
\vspace{0.1mm}
\bibliography{hirano}
\end{document}